# There are no valid points of criticism in Tyshkovskiy and Panchin's response (10.1002/bies.202000325) to our paper "The genetic structure of SARS-CoV-2 does not rule out a laboratory origin" (DOI: 10.1002/bies.202000240)

**Yuri Deigin[1] and Rossana Segreto[2]**

[1]Youthereum Genetics Inc., Toronto, ON Canada.

[2]Department of Microbiology, University of Innsbruck, Austria.

No external funding was received for this work.

The two authors contributed equally to the manuscript.

## Abstract

Tyshkovskiy and Panchin have recently published a commentary on our paper in which they outline several "points of disagreement with the Segreto/Deigin hypothesis". As our paper is titled "The genetic structure of SARS-CoV-2 does not rule out a laboratory origin", points of disagreement should provide evidence that rules out a laboratory origin. However, Tyshkovskiy and Panchin provide no such evidence and instead attempt to criticize our arguments that highlight aspects of SARS-CoV-2 that could be consistent with the lab leak hypothesis. Strikingly, Tyshkovskiy and Panchin's main point of criticism is based on a false premise that we have claimed RaTG13 to be a direct progenitor of SARS-CoV-2, and their other points of criticism are either incorrect or irrelevant to our hypotheses. Thus, the genetic structure of SARS-CoV-2 remains consistent with both natural or laboratory origin, which means that both the zoonotic and the lab leak hypothesis need to be investigated equally thoroughly.

## Introduction

In our paper "The genetic structure of SARS-CoV-2 does not rule out a genetic manipulation"[1] we describe several features of SARS-CoV-2 which could be consistent with a lab origin to demonstrate that the lab leak hypothesis cannot be dismissed as a conspiracy and should be investigated on par with a possible natural origin. Tyshkovskiy and Panchin have recently published a commentary in BioEssays[2] that attempts to criticize our paper and tries to dismiss the hypothesis that SARS-CoV-2 might have escaped from a laboratory. Herein we demonstrate that their commentary is built on the false premise that we have suggested RaTG13 to be a direct progenitor of SARS-CoV-2, and that other points of criticisms expressed by the authors are either irrelevant or not valid.

## A false premise: RaTG13 was not suggested as a direct progenitor of SARS-CoV-2

The main point of criticism of our paper[1] expressed by Tyshkovskiy and Panchin in their commentary recently published by BioEssays[2] is based on a false premise: "the first major problem with Segreto's and Deigin's hypothesis is the significant divergence between the genome sequence of SARS-CoV-2 and its proposed ancestor RaTG13". The quoted statement is clearly false because we never claim that RaTG13 itself is a "proposed ancestor" or the backbone used for a possible construction of SARS-CoV-2. The closest thesis we ever put forth is in the context of the observations that while RaTG13 is 96% close to SARS-CoV-2, its receptor binding domain (RBD) is highly divergent, whereas pangolin CoV MP789 shares a nearly identical RBD with SARS-CoV-2 at the amino acid level:

*Although its overall genome similarity is lower to SARS-CoV-2 than that of RaTG13, the MP789 pangolin strain isolated from GD pangolins has an almost identical RBD to that of SARS-CoV-2. Indeed, pangolin CoVs and SARS-CoV-2 possess identical amino acids at the five critical residues of the RBD, whereas RaTG13 only shares one amino acid with SARS-CoV-2.[35] ACE2 sequence similarity is higher between humans and pangolins than between humans and bats. Intriguingly, the spike protein*

*of SARS-CoV-2 has a higher predicted binding affinity to human ACE2 receptor than to that of pangolins and bats. Before the SARS-CoV-2 outbreak, pangolins were the only mammals other than bats documented to carry and be infected by SARS-CoV-2 related CoV. Recombination events between the RBD of CoV from pangolins and RaTG13-like backbone could have produced SARS-CoV-2 as chimeric strain. For such recombination to occur naturally, the two viruses must have infected the same cell in the same organism simultaneously, a rather improbable event considering the low population density of pangolins and the scarce presence of CoVs in their natural populations. Moreover, receptor binding studies of reconstituted RaTG13 showed that it does not bind to pangolin ACE2.*

Thus, to explain this observation we hypothesize that some sort of recombination might have occurred between a RaTG13-**like** backbone and an RBD from a MP789-**like** pangolin CoV. Moreover, we do not rule out either a natural recombination, or an engineered one.

Therefore, the stated claim of Tyshkovskiy and Panchin about our hypothesis' major problem is simply false. Consequentially, the authors' subsequent building on their false claim is irrelevant to our paper. Noticeably, however, even that analysis is flawed, as the authors seem to imply that their observation that the pattern of nucleotide substitution between SARS-CoV and Rs4231 matches the pattern between SARS-CoV-2 and RaTG13 is somehow indicative of a natural origin of SARS-CoV-2. That is not the case, as the mutation patterns could be the same even if SARS-CoV-2 was a product of lab-induced mutations (as a result of passaging in lab animals or cell cultures), i.e. the observation pointed out by the authors cannot be used to rule out a lab origin or prove a natural one. Moreover, the authors' claim that synonymous mutations cannot be generated by site-directed mutagenesis is false, as Cuevas et al.[3] have generated 53 single random synonymous substitution mutants of two bacteriophages by site-directed mutagenesis to assay their fitness.

## A weak point of criticism: a bat origin of SARS-CoV-2 with direct spillover to humans is highly unlikely

As a third point, the authors claim that under the natural origin hypothesis, the recombination events possibly leading to the emergence of SARS-CoV-2 did not need to occur in a pangolin, which we describe as an unlikely host for CoV recombination, but could instead have happened directly in bats, because "many related strains of the pangolin coronavirus have been discovered in bats". The quoted statement is vacuous and irrelevant because bats are the primary hosts of SARS-related CoVs, and thus it is pangolins, not bats, where one should be surprised to find a SARS-like CoV. To our knowledge, the pangolin strains GD2019 and GX2017 found in the pangolins seized by Guangxi customs are the only such known strains. Moreover, when South Asian pangolins entering the wild life trade were recently tested for the presence of CoVs, none turned up positive.[4] Notably, the pangolin CoV MP789 is the only known CoV to have an RBD almost identical at the amino acid level to the one found in SARS-CoV-2. This RBD is peculiar because it is characterized by a very high binding affinity to the human ACE2 receptor, but it binds poorly to the bat ACE2 receptor,[5, 6] making unlikely the authors' suggestion of a bat origin of SARS-CoV-2. This observation also negates the authors' suggestion that SARS-CoV-2's spike protein is not chimeric and that its RBD must be ancestral, but instead it is RaTG13 that is chimeric and whose RBD is a result of recombination.

## Tyshkovskiy and Panchin's analysis of restriction sites in and around SARS-CoV-2's furin cleavage site is flawed

As a fourth point, the authors dismiss as unsurprising our observation that the 12-nucleotide insertion that has created a polybasic furin cleavage site (FCS) in SARS-CoV-2 contains a *Fau*I restriction enzyme site conveniently positioned on the nucleotides coding for the two newly inserted arginines, which enables using the restriction fragment length polymorphism (RFLP) technique for quick colony screening to check whether the FCS – which is prone to deletion *in vitro* – is still present. The authors posit that their analysis of hundreds of different restriction enzymes makes the existence of ***some*** restriction enzyme site in any 12-nucleotide stretch a virtual certainty, ~99.5%.

However, in their calculation, the authors mistakenly assume that the probability of any single nucleotide being a part of some restriction enzyme's recognition site is independent of other nucleotides, which is clearly not the case because recognition sites span four or more consecutive nucleotides. Moreover, many recognition sites tend to occur in clusters because they share some part of the recognition site motif. The authors' own Fig. 3 is a great illustration of this: there are some clusters of restriction enzyme hotspots, and there also are large stretches of consecutive nucleotides without any restriction enzyme sites – many such stretches significantly longer than 12 nucleotides in length – thereby providing a simple visual refutation of the authors' own 99.5% estimate.

Also, the authors have included 4-nucleotide recognition site enzymes in their calculation, of which in a 30000-nucleotide genome there would be dozens if not hundreds for each such enzyme (Fig. 1), thereby heavily biasing the outcome of their calculation. The use of 4-nucleotide recognition site restriction enzymes which are frequent cutters would result in a hardly discernible pattern with no benefit for using the RFLP technique. In contrast, *Fau*I is an enzyme with a 5-nucleotide recognition site and there are only a handful of *Fau*I sites in any known coronavirus. SARS-CoV-2 has just six such sites, including one in the newly inserted FCS and having a new *Fau*I site show up in a 12-nucleotide insertion is an event worthy of notice.

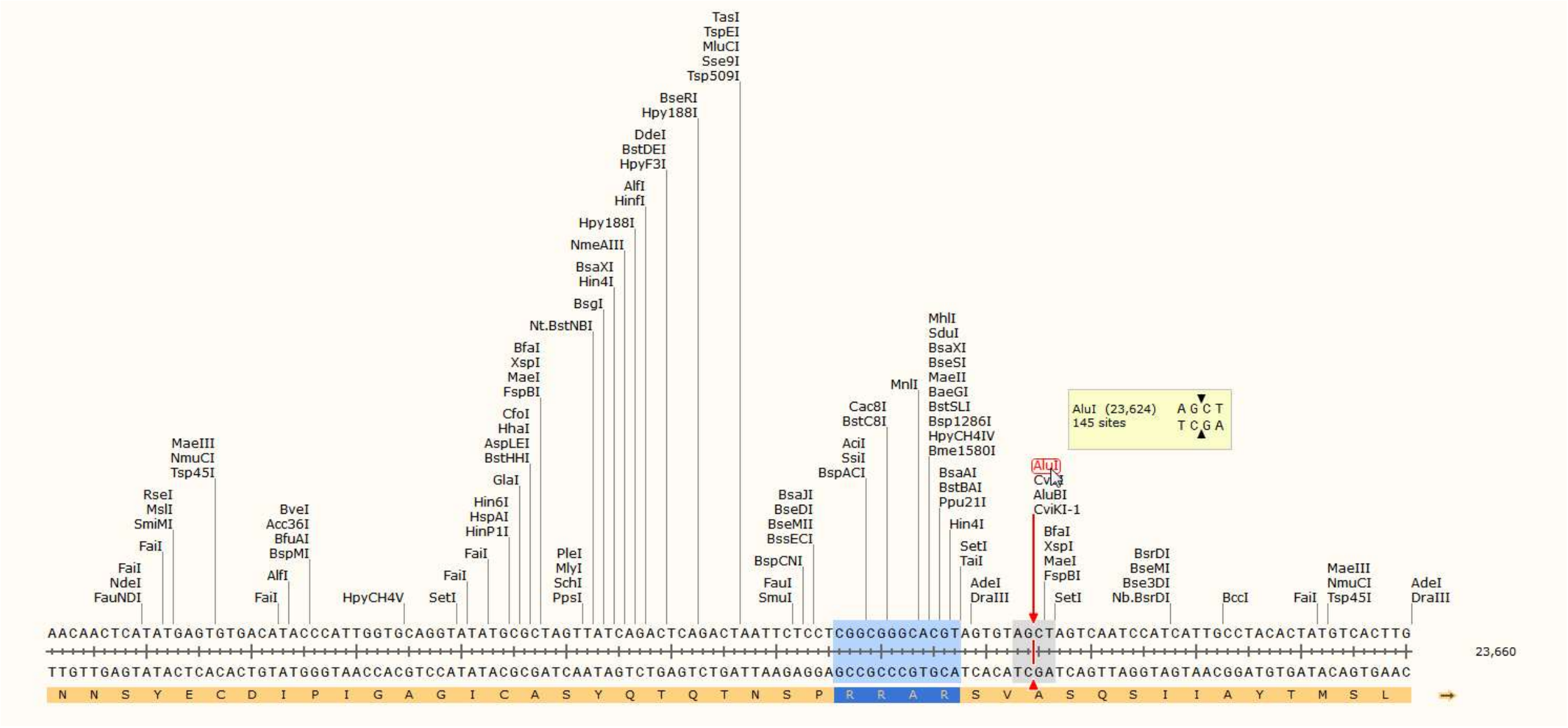

*Figure 1. Tyshkovskiy and Panchin included restriction enzymes with a 4-nucleotide recognition site in their analysis of probability of finding an enzyme recognition site within any given 12-nucleotide stretch of a coronavirus genome, thereby obviously biasing the odds of finding such sites. However, FauI has a 5-nucleotide recognition site, and there are only six such sites in all of SARS-CoV-2's 30k genome.*

Finally, the authors fail to appreciate that the *Fau*I site is notable for its unique property which enables it to be used to screen precisely whether the two arginines (R) in the newly created RRAR polybasic cleavage site are still present, as *Fau*I's recognition sequence is created by the CGGCGG codons coding for RR. What is even more notable is that CGG codons are the rarest codons to code for arginine in SARS-CoV-2 and its related CoVs (Fig. 2), unlike in humans where they are the most frequent,[a] so to see two CGGs inserted in a row is another low probability event. The authors also mention that they are not aware of the use of *Fau*I restriction site to screen mutations in coronavirus, without considering that the sequence coding for the FCS in SARS-CoV-2 is not present in any other coronavirus identified so far, and that the RFLP technique can be tailored to any restriction enzyme.

---

[a] GenScript Codon Usage Frequency Table (chart) Tool. https://www.genscript.com/tools/codon-frequency-table (last accessed on Mar 18, 2021).

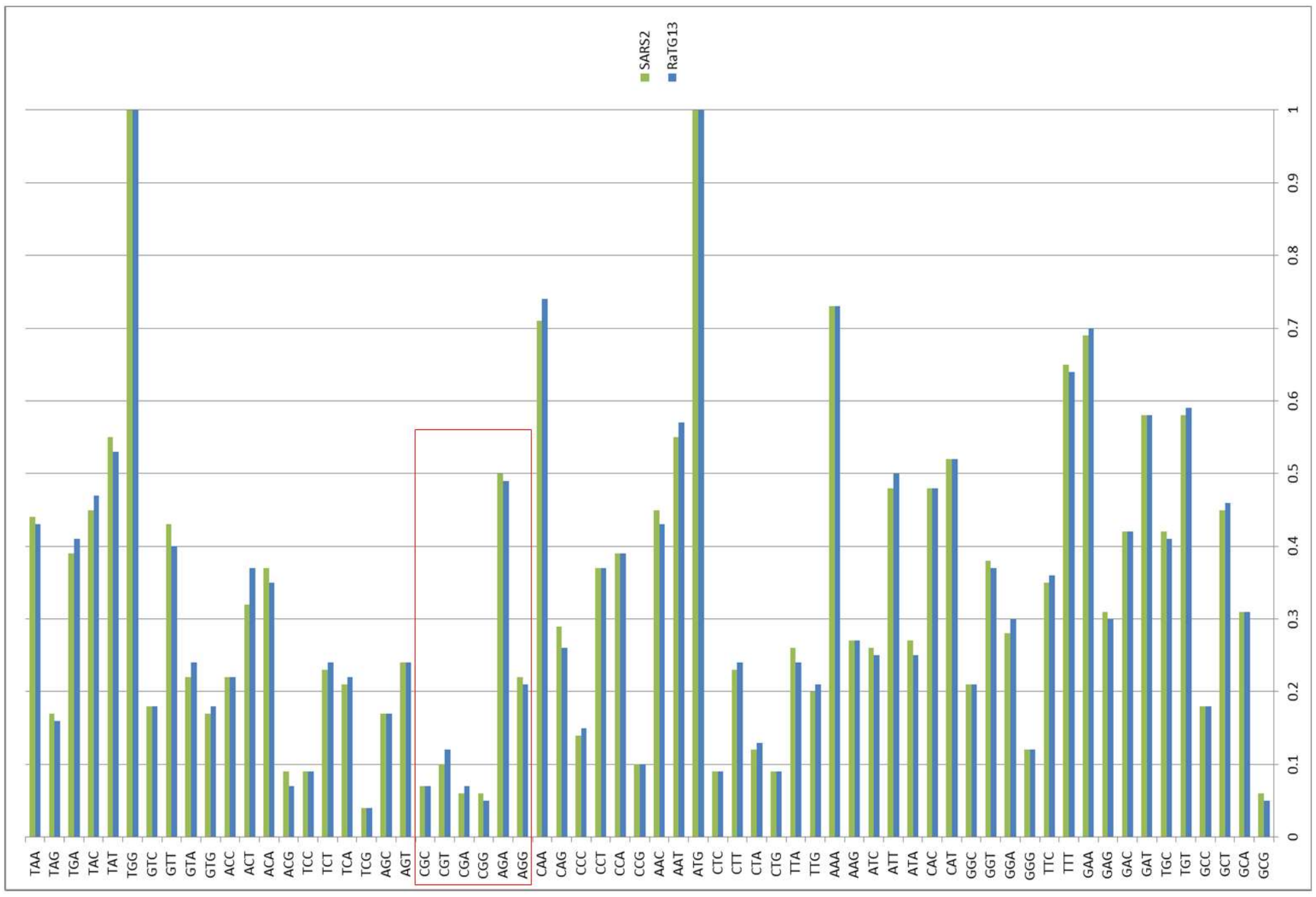

*Figure 2. Codon usage preference of SARS-CoV-2 and RaTG13 (codons coding for arginine are marked by a red box)*

## The out-of-frame insertion of the furin cleavage site is not proven to be natural by Tyshkovskiy and Panchin's cited papers

In their fifth point, Tyshkovskiy and Panchin criticize our observation that the insertion of the sequence coding for the FCS in SARS-CoV-2 is not in frame with the rest of the sequence and theorize that the reading frame of the spike protein is maintained because the sequence is a multiple of three nucleotides (12 in total). It should be noted that the codon (CCT) used for proline (P) in the inserted PRRA sequence (T**CCT**CGGCGGGC) is preceded by the single nucleotide "T" and the reading frame is therefore maintained because the insertion of the FCS sequence generates a split of the codon for serine (S), as we describe in our article. The authors also refer to Perdue et al.[7] to demonstrate that such insertions are commons in influenza viruses, but the article they cite describes sequence duplication of the cleavage site which has resulted neither in a frame shift nor in a codon split. The same article postulates that only insertions of multiples of three are genetically stable, confirming the peculiarity of the FCS insertion in SARS-CoV-2 with a single nucleotide preceding the FCS sequence.

## Conclusion

In closing, Tyshkovskiy and Panchin' attempted use of the principle of maximum parsimony in the context of deciding whether a natural origin is more likely than a lab leak is outright amusing, especially considering that the authors state in their introduction that a lab leak can only be established by investigating the lab in question and not "by analyzing the genetic and phenotypic properties of the virus". This not only seems to contradict the title of their commentary but also greatly diminishes the entire point of their work. There clearly are several observations about SARS-CoV-2 and the research that has been ongoing in Wuhan that are consistent with both natural and lab escape origin of the virus. Moreover, the authors are mistaken that genetic or phenotypic properties of the virus cannot potentially provide evidence of lab origin. In fact, it was precisely the phenotypic characteristics consistent with vaccine development, namely temperature sensitivity, which have ultimately established the scientific consensus that the 1977 H1N1 flu pandemic was due to a lab leak.[8]

In conclusion, not a single point of criticism expressed by Tyshkovskiy and Panchin about our article is valid and their comment is based on false assumptions and consists of analysis mostly irrelevant to our hypotheses.